\documentclass[preprint2]{aastex6}

\usepackage{lineno}
\usepackage{graphicx}

\usepackage{xcolor}

\usepackage[utf8]{inputenc}

\newcommand{\msun}{\ensuremath{M_\odot}}
\newcommand{\msunyr}{\ensuremath{\msun\,\mathrm{yr}^{-1}}}
\newcommand{\mdot}{\ensuremath{\dot{M}}}
\newcommand{\ha}{\ensuremath{\mathrm{H}\alpha}}
\newcommand{\PV}{P\,{\sc v}\,\ensuremath{\lambda \lambda 1118, 1128}}

\begin{document}

\title{Ultraviolet Spectropolarimetry with Polstar: Clumping and Mass-loss Rate Corrections}

\author{Ken Gayley}
\affil{Department of Physics and Astronomy, University of Iowa, Iowa City, IA, 52242}
\author{Jorick S. Vink}
\affil{Armagh Observatory and Planetarium, College Hill, BT65 9DG Armagh, Northern Ireland}
\author{Asif ud-Doula}
\affil {Penn State Scranton, 120 Ridge View Drive, Dunmore, PA 18512, USA}
\author{Alexandre David-Uraz}
\affil{Department of Physics and Astronomy, Howard University, Washington, DC 20059, USA\\
Center for Research and Exploration in Space Science and Technology, and X-ray Astrophysics Laboratory, NASA/GSFC, Greenbelt, MD 20771, USA}
\author{Richard Ignace}
\affil{Department of Physics \& Astronomy,
East Tennessee State University,
Johnson City, TN 37614, USA}
\author{Raman Prinja}
\affil{Department of Physics and Astronomy, University College London,
Gower Street, London WC1E 6BT, UK}
\author{Nicole St-Louis}
\affil{D\'epartement de physique, Universit\'e de Montr\'eal, Complexe des Sciences, 1375 Avenue Th\'er\`ese-Lavoie-Roux, Montr\'eal (Qc), H2V 0B3, Canada}
\author{Sylvia Ekstr\"om}
\affil{Department of Astronomy, University of Geneva, Chemin Pegasi 51, 1290 Versoix GE, Switzerland}
\author{Ya\"el Naz\'e}
\affil{GAPHE, Univ. of Li\`ege, B5C, All\'ee du 6 Ao\^ut 19c, B-4000 Li\`ege, Belgium}
\author{Tomer Shenar} 
\affil{Anton Pannekoek Institute for Astronomy and Astrophysics, University of Amsterdam, 1090 GE Amsterdam, The Netherlands}
\author{Paul A. Scowen}
\affiliation{NASA GSFC}
\author{Natallia Sudnik} 
\affil{Nicolaus Copernicus Astronomical Centre
of the Polish Academy of Sciences, Bartycka 18, 00-716 Warsaw, Poland}
\author{Stan P. Owocki}
\affil{Department of Physics and Astronomy, University of Delaware, 217 Sharp Lab, Newark, DE, USA}
\author{Jon O. Sundqvist}
\affil{Institute of Astronomy, KU Leuven, Celestijnenlaan 200D/2401, 3001 Leuven, Belgium}
\author{Florian A. Driessen}
\affil{Institute of Astronomy, KU Leuven, Celestijnenlaan 200D/2401, 3001 Leuven, Belgium}
\author{Levin Hennicker}
\affil{Institute of Astronomy, KU Leuven, Celestijnenlaan 200D/2401, 3001 Leuven, Belgium}

\begin{abstract}
The most massive stars are thought to lose a significant fraction of their mass in a steady wind during the main-sequence and blue supergiant phases.  This in turn sets the stage for their further evolution and eventual supernova, with consequences for ISM energization and chemical enrichment. Understanding these processes requires accurate observational constraints on the mass-loss rates of the most luminous stars, which can also be used to test theories of stellar wind generation.  In the past, mass-loss rates have been characterized via collisional emission processes such as H$\alpha$ and free-free radio emission, but these so-called ``density squared'' diagnostics require correction in the presence of widespread clumping. Recent observational and theoretical evidence points to the likelihood of a ubiquitously high level of such clumping in hot-star winds, but quantifying its effects requires a deeper understanding of the complex dynamics of radiatively driven winds.  Furthermore, large-scale structures arising from surface anisotropies and propagating throughout the wind can further complicate the picture by introducing further density enhancements, affecting mass-loss diagnostics. Time series spectroscopy of UV resonance lines with high resolution and high signal-to-noise are required to better understand this complex dynamics, and help correct ``density squared'' diagnostics of mass-loss rates. The proposed 
\textit{Polstar} mission easily provides the necessary resolution at the sound-speed scale of 20 km s$^{-1}$, on three dozen bright targets with signal-to noise an order of magnitude above that of the celebrated IUE MEGA campaign, via continuous observations that track structures advecting through the wind in real time.

\end{abstract}

\maketitle
\section{Introduction}
%
%{\color{blue}I added a mention of this white paper in S3's (on the origin of rapidly rotating B stars) introduction. If you want any information %about our whitepaper to link to yours just let me (Carol Jones) know.}
Theoretical models of the UV line-driven winds of hot stars \citep{1988ApJ...335..914O} %(Owocki, Castor \& Rybicki 1988) 
predict that internal instabilities should produce
high-density clumps and low-density voids, as the hypersonic acceleration cannot be maintained smoothly.
Also, in CAK theory \citep*{1975ApJ...195..157C}, %(Castor, Abbott \& Klein ) 
the acceleration of the wind depends nonlinearly on the mass flux, so variations
in the mass flux launched from the stellar photosphere also produce significant velocity and density variations as the wind accelerates.
Both of these types of wind perturbations will steepen into shocks, 
compressing and driving dense clumps \citep{2018A&A...611A..17S}, %Sundqvist, Owocki \& Puls 2018
and producing observable hard X-ray emission (e.g., Oskinova 2016 and references therein). %\citep{2016AdSpR..58..679O}[and references therein]

Direct observational evidence that this is occurring even in single stars already exists in optical emission 
lines of dense winds, in moving emission bumps
\citep[e.g.,][]{1988ApJ...334.1038M, 1999ApJ...514..909L} and reduced levels of free-electron scattering in the line wings \citep{1991A&A...247..455H, 1998A&A...335.1003H}.%(Hillier 1991)
For all these reasons, the concept of a constant mass-flux, smoothly accelerating hypersonic flow is likely not realized in actual winds.

Nevertheless, most current empirical analyses of stellar atmospheres and winds rely on the use of spherical non-LTE
model atmospheres. In state-of-art analyses the stellar and wind parameters, such as 
$\dot{M}$, are derived by fitting resonance and recombination lines simultaneously.
P\,Cygni lines (often high opacity lines such as C\,{\sc iv}) are predominately used to determine terminal wind velocities $v_{\infty}$ 
(E.g. \citealt{1990ApJ...361..607P}%Prinja et al. 1990
), whereas \mdot\ in the stronger winds of O-stars 
(mass-loss rates above $10^{-7}$ \msunyr\ \citep[e.g.][]{2007A&A...465.1003M}) % Mokiem+ 2007 
is often inferred using \ha emission.
The latter benefits from arising from what is incontrovertibly the dominant wind 
species in hot main-sequence stars, ionized H \citep[e.g.][]{1996A&A...305..171P}.%Puls et al. 1996

Yet there is a notable difference between UV resonance lines 
and (optical) recombination lines such as H$\alpha$ (see left-hand side vs  right-hand side of Fig.\,\ref{f_oski}). 
The resonance lines involve scattering of pre-existing UV stellar continuum, and their optical
depth $\propto \rho$, while recombination involves photon creation by collision of 
2 particles, such that line emissivities $\propto \rho^2$. 
The latter also holds for the free-free radio continuum, so crucial indicators of global wind mass-loss rates
are sensitive to the local density variations induced by clumping.
Thus, use of these ``density squared'' diagnostics allow 
inhomogeneities to mimic a higher mass-loss rate.
Wind inhomogeneities can also induce strong variations in the local velocity gradients (an effect
termed ``vorosity'' in Sundqvist et al. 2014), which allow more of the stellar continuum to penetrate through the
wind opacity of otherwise saturated UV resonance lines, which can mimic a \textit{lower} mass-loss rate
in absorption-line diagnostics.
Hence, the dual analysis of UV plus optical lines is affected by clumping
in subtle and important ways that must be informed by new types of observations.

This has significance not only for our understanding of the wind dynamics themselves, but also for establishing the evolutionary
consequences of mass loss during phases of continuous wind outflow, especially the main sequence.
The latter is the longest-lived phase of stellar evolution, and happens first, establishing the initial conditions for
all subsequent evolutionary phases-- including the ultimate supernova of the massive star.
Furthermore, the high-mass end of the main sequence has the most profound impact on galactic evolution because it reprocesses the stellar
material on such short timescales, all while lower-mass stars are still in their formative stages.
Thus, to understand the evolution of massive stars and their impact on the host galaxy in population synthesis studies,
we must first accurately determine the mass lost in these decisive initial phases (Langer 2012), %citep{2012ARA&A..50..107L}
and track how it evolves in the blue supergiant phase.
This is especially true for the highest-mass stars, whose wind mass-loss rates are highest and have the greatest
evolutionary significance for all subsequent phases of the star, but such stars are also the rarest.
Somewhat less massive stars
afford more numerous examples to study, and the lessons learned from them place our understanding into a
fuller context, so we wish to see the impact of clumping over the fullest possible range of massive stars.

Yet despite the this need for clumping corrections to
observationally inferred mass-loss rates,
no systematic ultra-high SNR study of clumping and CIR-type features in dynamical spectra of massive star winds has yet
been undertaken.
The IUE MEGA campaign \citep{1995ApJ...452L..53M} 
was only able to obtain SNR ~ 20-25, and only on a few targets.
The efforts by HST were also limited
to a few targets, and were hampered in some cases by pointing difficulties, and an orbit that only allowed continuous observation for 
about an hour at a time. % ( zeta Oph).  
Hence these past experiments were only able to hint at the presence of clumping and dynamics on scales smaller than the
gross ``discrete absorption component'' (DAC) structures that are often seen 
in massive-star winds \citep{1997A&A...327..281K}, 
yet these structures
are believed to be responsible for extreme density variations necessary to account for the filling-in of P V absorption \citep{2006ApJ...637.1025F}
and the absence of free-electron scattering wings 
in dense winds \citep{1991A&A...247..455H}.

Enter \textit{Polstar}, whose orbit allows continuous coverage for days, at wavelength resolution $R = 33,000$ in Ch1.
One objective (termed \textit{S2}) of the \textit{Polstar} mission is deep, continuous exposures of the ~40 brightest
massive stars over a range of masses and OB spectral types, to watch clumping and other structure develop in real time in their winds.
The fundamental physics of the wind driving is informed by the nature of this clumping, and larger structures such as co-rotating
interactions regions (CIRs, c.f. Cranmer \& Owocki 1996), %\citep{1996ApJ...462..469C}
as well as prolate/oblate asphericities.
But the primary goal of \textit{S2} is to make the necessary corrections to clumping-sensitive mass-loss rate determinations, such as free-free radio  
(Lamers \& Leitherer 1993) and H alpha emission (Puls et al. 1996), because of its importance to stellar evolution as described next.
% Puls, J., Kudritzki, R.-P., Herrero, A., Pauldrach, A. W. A., Haser, S. M., Lennon, D. J., 
% Gabler, R. ; Voels, S. A. ; Vilchez, J. M. ; Wachter, S. ; Feldmeier, A.
% Journal: Astronomy and Astrophysics, v.305, p.171

\section{The importance of accurate mass-loss rate determinations}

\begin{figure}[h!]
    \centering
    \includegraphics[width=7cm]{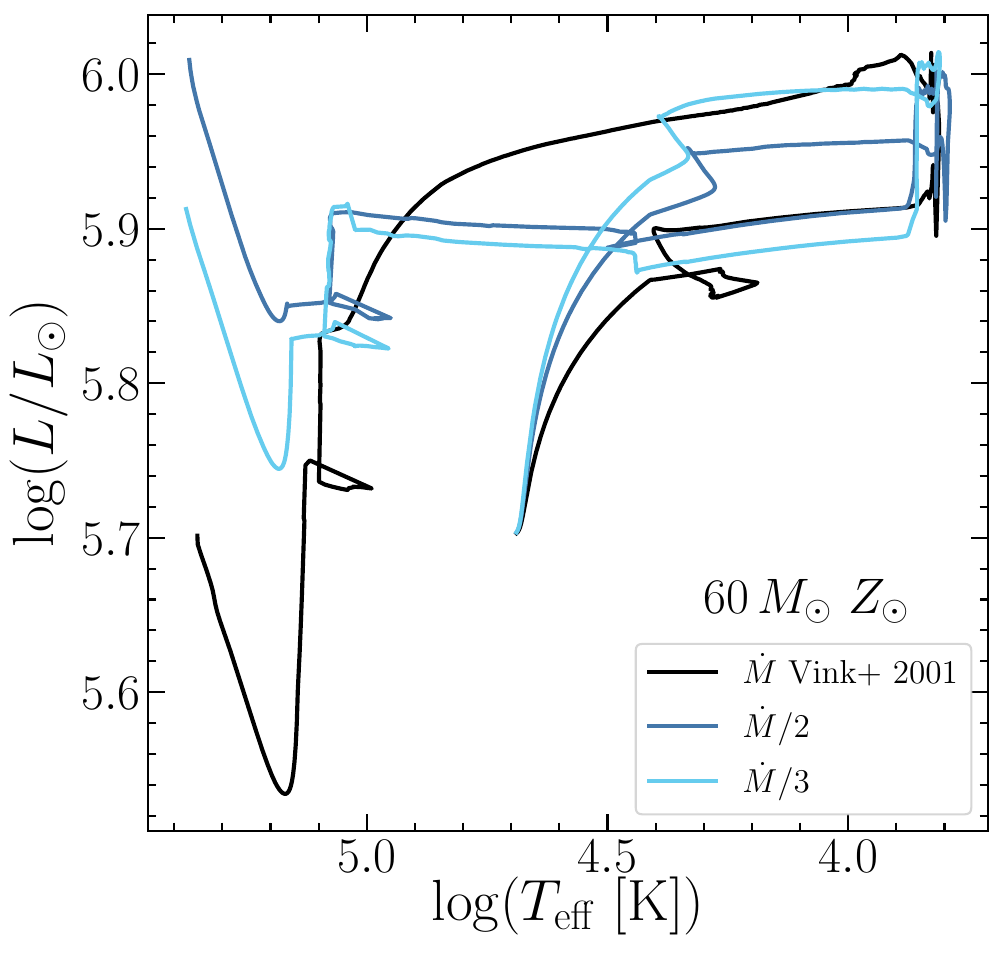}
    \caption{Hertzsprung-Russell diagram of 60\,\msun\ models computed with mass-loss recipe of \citet[black]{2001A&A...369..574V}, % Vink+ 2001
    and the same recipe divided by a factor of 2 and 3 during the MS (teal blue and cyan, respectively).}
    \label{fig:compaMdot}
\end{figure}

Models of the most massive stars confirm the observational implication that mass loss can play
a crucial role in stellar evolution, and accurate knowledge of the
mass-loss rate is required for understanding the impact on the host galaxy.
Figure~\ref{fig:compaMdot} presents the evolution of 60\,\msun\ models computed until the end of central carbon burning, and shows how this evolution is altered when the mass-loss rates \citep[chosen from][]{2001A&A...369..574V} are reduced by a factor of 2 or 3 (here the factors have been applied only during the main sequence, MS). The difference is mainly due to larger winds during the MS (in the standard case) reducing the size of the core, and hence reducing the overall luminosity: at the middle of the MS, the \mdot/2 model has a core that is 3\% larger than the standard model, and the \mdot/3 has a core that is 5\% larger. In the middle of central helium burning, the difference increases, amounting to 50\% and 45\% respectively. At the end of central C-burning, the CO-core mass ($M_\mathrm{CO}$) is 57\% (35\%) larger in the \mdot/2 (\mdot/3) models respectively, leading to very different endpoint locations.
A direct effect of changing the mass-loss rates is the duration of the Wolf-Rayet (WR) phase expected for the models. The standard model becomes a WR early in the He-burning phase ($X(^4\mathrm{He})=0.97$), while the lower \mdot-rates models arrive later in this regime ($X(^4\mathrm{He})=0.29$ and $0.44$ respectively). This affects directly their statistical observability.

Surprisingly, the models show that the behaviour is not simply monotonic with the reduction factor on the mass-loss rates,
revealing the sensitivity of the advanced phases to what precedes them. 
In stars, gravity and fusion interact in complex ways, and changing one parameter in the modeling feeds back on all other aspects. Because of its high luminosity at the end of the MS, the \mdot/3 model loses more mass during the crossing of the Hertzsprung-Russell diagram, spends less time with a $\log(T_\mathrm{eff})<4.0$, and ends its evolution in an intermediate situation between the standard and the \mdot/2 models.
Hence, changes to mass-loss rates at the factor 2 level can profoundly alter the initial conditions for all subsequent evolution, ultimately
changing the final state where supernova occurs.
Since internal structure at all scales can induce such changes in the mass-loss rates inferred from H $\alpha$ and radio emissions,
understanding high-mass stellar evolution requires resolving and analyzing the dynamical nature of wind clumping.

\section{The \textit{Polstar} experiment for objective \textit{S2}}

The effective area ($A$ = 50 cm$^2$ at 150 nm) and spectral resolution ($R = 33,000$) of channel 1 (hereafter Ch1) of the
\textit{Polstar} instrument are ideally suited for continuous observing of bright targets that can track features
accelerating through the line profile in real time.
The spectral resolution ($\cong$ 10 km s$^{-1}$) penetrates to the Sobolev scale of thermal Doppler shifts in the metal-ion
UV resonance lines observed, and the area $A$ provides sufficient signal-to-noise (SNR) for exposures of duration
\begin{equation}
    t \ = \ \frac{c}{aR} \ ,
\end{equation}
where $a$ is the characteristic acceleration rate of the features.
The correct value of $a$ is currently unknown, and may vary from star to star, but we might expect
it to be of order the stellar surface gravity $g$, which is typically $\sim 0.03$ km s$^{-2}$.
So we will synchronize the exposure time with the rate that features move across resolution elements, 
and taking a canonical acceleration of $a \sim 0.03$ km s$^{-2}$ 
% (inferred from Massa \& Prinja 2015) % \citep{2015ApJ...809...12M} 
implies $t \cong 400 s$.
% Since structure formation below the sound speed will be inhibited by gas pressure, we need only resolve $\cong$ 20 km s$^{-1}$
% (the sound speed at $T \sim 30,000 K$),
% so binning in wavelength can yield even higher SNR if desired.
Given \textit{Polstar's} effective area, $t \cong 400 s$ and $R \cong 30,000$ implies the SNR is
\begin{equation}
    \frac{S}{N} \ \cong \ 90 \sqrt{f} \ ,
\end{equation}
where $f$ is the target flux in units of $10^{-9}$ erg cm$^{-2}$ s$^{-1}$ nm$^{-1}$, 
a characteristic 
flux for an average \textit{Polstar} target but somewhat dim for this experiment.
Thus dynamical spectra can be produced by continuous monitoring for days, as was done with the IUE MEGA campaign 
% (Massa \& Prinja 2015), 
\citep{2015ApJ...809...12M} ,
with greatly improved SNR ($S/N > 400$) for bright targets ($f > 20$).

There are 44 potential targets in our list that satisfy this brightness limit (shown in Table 1).
Because of their
special importance, we added 2 additional sources ($\zeta$ Oph and $\phi$ Per)
that are within a factor of 2 of the brightness limit, for a total of 46
potential targets for the \textit{S2} objective.
The mission would select $\sim$ 40 of these for continuous observation over
a day or so, based
on flexible criteria such as accessibility in the scheduling and overlap with other objectives.

For example, 15 of the targets with $f > 25$ will be prioritized because
they also appear in the target list for objective
\textit{S3},  
relating to rapid rotation
causes \citep{2021arXiv211107926J}, which we can connect with clumping effects.
The star $\zeta$ Oph was observed by HST for under 5 hours, but it provides a useful comparison with previous results,
and $\phi$ Per is a variable system involving spinup by binary interaction, allowing the effects of 
rapid rotation on wind dynamics to be studied.
Both added stars will produce SNR $>$ 250, or an order of magnitude above the IUE MEGA campaign, while all the rest
exceed SNR $\cong$ 400.
This sensitivity is deemed necessary to penetrate the full range of the dynamical structures down to the Sobolev scale $ \cong R/100$,
since at that scale, we expect $ \cong 100^2$ cross sections of Sobolev size across the face of the star, with a statistical
variance of some 1 \%, requiring SNR at the several-hundred level to confidently resolve.

Once the features are detected, the
exposure time ($\cong$ 400 s) is matched to the 
velocity resolution ($\cong$ 10 km s$^{-1}$) to allow the features to be tracked
as they accelerate through the wind, to reveal how they initiate and evolve, 
and quantify their impact
on clumping-sensitive mass-loss determinations.
And if the acceleration of the features should prove faster than the canonically chosen $a \sim 0.03$ km s$^{-2}$, even higher
SNR will be achieved by simply binning in wavelength.
If the features cannot then be resolved at the Sobolev scale ($\sim 10$ km s$^{-1}$),
they can still be resolved at the sonic scale ($\sim 20$ km s$^{-1}$).
% A similarly small degree of binning can also be used to extend high SNR $> 400$ to the 40 brightest % targets if desired.

The number of targets in the current plan, $\sim$ 40, 
can be monitored for a day or so, or longer in special cases (such as stars like $\zeta$ Pup with rotation periods
fast enough to study periodicity), within
the observing time set aside for this objective (\textit{S2}), as part of the 3-year science mission.
This is an opportune number of targets, because it is sufficient to consider trends with mass and spectral type, and it
also explores the ubiquity of DACs, and smaller-scale clumps, stemming from either surface perturbations or the LDI.

The other objectives can also benefit from this wealth of observational data on the three dozen brightest massive
stars in the sky, as the data can be stacked and binned to achieve unprecedented UV polarization precision.
For example, for a median star from the list like $\gamma$ Cas, stacking over 2-hour chunks to allow temporal
variations over the observation to be studied, and binning to $R = 30$ to focus on continuum polarization, would allow
a SNR of some $1 \times 10^5$ per resolution element.
Even with the lower effective area of Ch1, that would support a polarization precision of better than
$3 \times 10^{-4}$ , in the same dataset that would already by mined for its dynamical spectroscopic potential.
The polarization information would reveal processes that break the spherical symmetry in a globally steady way,
while the dynamical spectra reveal evolving process on much smaller spatial and temporal scales within the wind.

The reason polarimetry is automatically obtained, simultaneously with the spectral information, is that \textit{Polstar}
determines all four Stokes parameters, $I$, $Q$, $U$, and $V$, in all its observations.
Many of the proposed experiments % [other white papers] 
will use the higher effective area of the low-resolution channel 2 (Ch2) when doing sensitive polarimetry,
but in this experiment, the continuous monitoring is in the high-resolution Ch1, so the polarimetric information will come 
from binning down to low resolution and simply accepting the roughly factor 3 loss of effective area in Ch1.

This loss is mitigated by the use of inherently bright targets, so when the SNR $\cong$ 500-1000 at $R = 33,000$, it will
be $\sqrt{1000}$ times better, or $\cong$ 15,000-30,000, at $R = 33$, a sufficient 
resolution for using the spectral shape to remove
foreground polarization from the interstellar medium 
citep{Andersson2021}.  % ISM white paper
Separation from background
polarization is also assisted by the time-dependent character of the 
intrinsic polarization from the wind dynamics.
Such high SNR is conducive to polarimetric precision at the required $3 \times 10^{-4}$ target, which is capable of
detecting the temporal variations from some 10,000 clumps with individual free-electron scattering optical depths of ~0.1.
%kgg:  we may have to consider dipole scattering components of UV resonance lines to get enough polarization signal 
Hence, the \textit{Polstar} mission objective \textit{S2} will combine dynamical spectra and time-dependent linear polarization
variations to track clump and CIR features down to the Sobolev scale, as they accelerate through the winds of ~40 bright
massive stars, to understand how clumping and structure form, and what is their impact on wind mass-loss rate determinations.
It can also detect global structures with very small latitudinal differences in optical depth, down below $\cong$ 0.01, 
due to rotation or other temporally persistent aspherical features, as well as rotationally modulated polarization
from large-scale azimuthal structures such as CIRs.

\begin{table}
\label{targettable}
\caption{{\em Polstar} Targets for Structured Wind Studies}
\begin{tabular}{llcc}
\hline\hline Target & Name & Flux (1500\AA) & SNR \\ 
 & & (10$^{-10}$ erg/s/cm$^2$/\AA) & eq. (2) \\ \hline
HD 122451 & $\beta$ Cen & 600 & 2200 \\
HD 108248 & $\alpha$ Cru & 600 & 2200 \\
HD 116058 & $\alpha$ Vir & 530 & 2000 \\
HD 11123 & $\beta$ Cru & 400 & 1800 \\
HD 68273 & $\gamma$ Vel & 260 & 1400 \\
HD 66811 & $\zeta$ Pup & 230 & 1300 \\
HD 37742 & & 200 & 1200 \\
HD 37742 & $\zeta$ Ori & 200 & 1200 \\
HD 116658 & Spica & 200 & 1200 \\
HD 35468 &  $\gamma$ Ori & 190 & 1200 \\
HD 52089 & $\epsilon$ CMa & 180 & 1200 \\
HD 44743 & $\beta$ CMa & 160 & 1100 \\
HD 36486 & $\delta$ Ori & 150 & 1100 \\
HD 24912 & & 130 & 1000 \\
HD 149438 & & 130 & 1000 \\
HD 5394 & $\gamma$ Cas & 100 & 900 \\
HD 205021 & $\eta$ Cep & 100 & 900 \\
HD 37043 & $\iota$ Ori & 100 & 900 \\
HD 34085 & Rigel & 90 & 800 \\
HD 127972 & $\eta$ Cen & 90 & 800 \\
HD 143018 & $\pi$ Sco & 75 & 700 \\
HD 151890 & $\mu^1$ Sco & 70 & 700 \\
HD 143275 & $\delta$ Sco & 70 & 700 \\
HD 105435 & $\delta$ Cen & 55 & 600 \\
HD 106490 & $\delta$ Cru & 55 & 600 \\
HD 87901 & Regulus & 50 & 600 \\
HD 120307 & $\nu$ Cen & 40 & 500 \\
HD 122451 & & 40 & 500 \\
HD 136298 & $\delta$ Lup & 37 & 500 \\
HD 157246 & $\gamma$ Ara & 35 & 500 \\
HD 205021 & $\beta$ Cep & 35 & 500 \\
HD 120324 & $\mu$ Cen & 34 & 500 \\
HD 147165 & $\sigma$ Sco & 32 & 500 \\
HD 11415 & $\epsilon$ Cas & 31 & 490 \\
HD 52089 & $\xi$ CMa & 30 & 490 \\
HD 121743 & $\phi$ Cen & 30 & 490  \\
HD 37202 & $\zeta$ Tau & 30 & 490 \\
HD 143118 & $\eta$ Lup & 30 & 490 \\
HD 50013 & $\kappa$ CMa & 29 & 480 \\
HD 19356 & Algol & 27 & 460 \\
HD 125238 & $\iota$ Lup &  27 &  460 \\
HD 3360 & $\zeta$ Cas & 27 & 460 \\
HD 10144 & & 27 & 460 \\
% HD 65816 &  & 20 & 400 \\
\hline
HD 10516 & $\phi$ Per & 15 & 350 \\  % this one is included because it is a variable near-critical rotator with importance to binary interactions
HD 149757 & $\zeta$ Oph &  10 &  280 \\  % this one is included because HST looked at it, we should get S/N = 100 at R = 33,000
\hline
\end{tabular}
\end{table}

\section{Small-scale clumpy wind structure}
\label{sec_clumping}

Due to the variability of spectral lines, as well as the presence of linear polarisation, astronomers have known for decades that stellar winds are not stationary but time-dependent, and that this leads to inhomogeneous, clumpy media \citep[see][]{2008A&ARv..16..209P},  
% 2008cihw.conf.....H).% Puls+ 2008, Hamann+ 2008
Motivated by 
hydrodynamic simulations \citep[e.g.,][]{1997A&A...322..878F, 2013MNRAS.428.1837S} including the line-deshadowing 
instability (LDI) \citep{1988ApJ...335..914O}, extreme density variations are expected to lead to large 
root-mean-square density ratios, $D = \sqrt{<\rho^2>}/<\rho>$, sometimes called a ``filling factor" (though it is not necessary,
nor physically realistic,
to regard the interclump medium as void).
For optical depths that depend linearly on density, such as free-electron optical depth, the mean 
opacity of a clumped medium is the same as that for a smooth wind, 
whereas opacities that scale with the square of density (such as for recombination lines like \ha),
the optical depths are enhanced by the clumping factor $D$.

The presence of large $D$ values due to extensive micro-clumping (clumps on size scales smaller
than a continuum mean-free-path, typical of most hot-star winds) yields an enhancement to mass-loss rates derived from
diagnostics that scale with the density squared, such as \ha\ emission.
That enhancement is a factor of $\sqrt{D}$, requiring a corresponding correction to
older mass-loss rates derived with the assumption of smooth winds \citep{2007A&A...465.1003M}.
However, determining the correct value of $D$ is difficult, and is the purpose of objective \textit{S2}.
The goal is to use the resonance line opacity that is liberally sprinkled throughout the UV, because it covers a wide
range from strong to weak lines, allowing access to wind regions both close to and far from the star, and over a range
of wind mass-loss rates.

One challenge with metal resonance lines is that the abundances and degree of ionization can be unknowns.
In the EUV, the \PV\ line has the advantage in certain O stars of being expected to be the dominant ionization stage,
so should in principle provide an accurate 
estimate of solely the mass-loss rate.
However, \citet{2006ApJ...637.1025F} %Fullerton et al. (2006) 
selected a large sample of O-stars, which also had 
$\rho^2$ (from \ha/radio) estimates available, and found that treating resonance lines as though 
they were linear in $\rho$ required
extreme clumping factors of up to $D \sim$ 400 (e.g. \citealt{2003ApJ...595.1182B}%Bouret et al. 2003
). 

But UV resonance lines are more complicated than that, because the Sobolev optical
depth in such lines depends not on density, but on the ratio of density to velocity gradient, $\rho (dv/dr)^{-1}$.
Since the local acceleration, and hence $dv/dr$, also depends on density,
the Sobolev optical depth scales much more steeply than linear in density.
For example, if we neglect time dependent terms in the force equation (a rough approximation), and also neglect any
feedback between density and line opacities (also imperfect), then at given velocity $v$ and radius $r$, the standard
treatment using the CAK $\alpha$ parameter gives
    \begin{equation}
        \frac{dv}{dr} \ \propto \ \rho^\alpha \left ( \frac{dv}{dr} \right )^{-\alpha} \ .
    \end{equation}
With $\alpha \cong 2/3$, we then have $dv/dr \cong \rho^{-2}$, which implies the Sobolev optical depth
in the UV resonance line in question obeys $\tau \propto \rho (dv/r)^{-1} \ \propto \ \rho^3$.
This is quite a steep dependence, owing to the dependence on velocity gradient, wherein spatial regions
of low density yield open holes in velocity space where there is low line optical depth, an effect termed
``vorosity'' (Sundqvist et al. 2018) in analogy with the ``porosity'' concept for spatial holes in
continuum opacity.
Hence clumping alters the degree of absorption in a P Cygni trough, because the vorosity allows a
window through the normally optically thick line \citep{2018A&A...611A..17S}, %(Sundqvist et al. 2018)
to a degree that we wish to observe and understand
in order to understand the necessary clumping corrections.

\subsection{Optically thick clumping (``macro''-clumping)} 
\label{sec_macro}

Some analysis of optically thick clumping effects in resonance lines has already been undertaken, taking into account
more than just the ``filling factor'' $D$, but also the distribution, size, and clump geometry.
The conventional description of macro-clumping is based on a clump size, $l$, and an average spacing of a statistical clump distribution, $L$, with porosity length $h=L^3/l^2$ \citep{2018MNRAS.475..814O}.  % Owocki \& Sundqvist 2018
This porosity length $h$ represents the key parameter defining a clumped medium, as it corresponds to the photon mean free path in a medium consisting of optically thick clumps.

\begin{figure*}
\begin{center}
\hspace{-0.2cm}
\includegraphics[width=\textwidth]{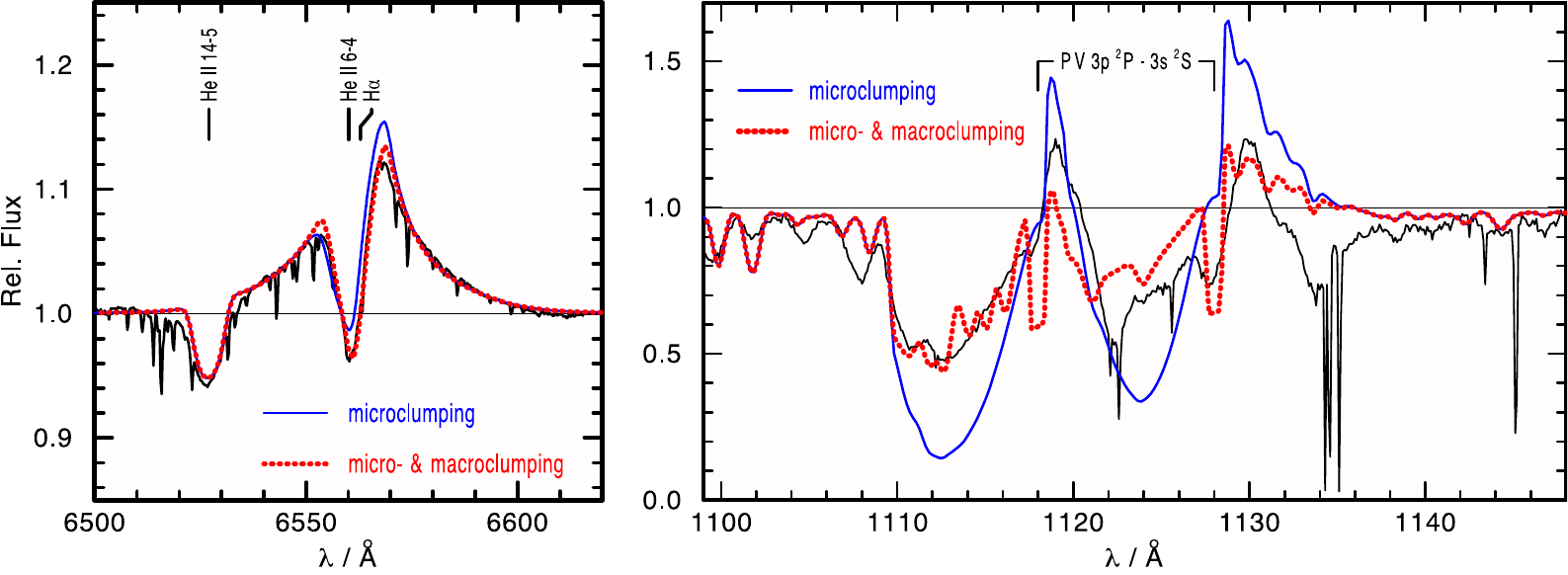}
\vspace{-0.2cm} 
\caption{Porosity as the likely solution for the PV problem. The \ha\ line on the left-hand side is hardly affected by macro-clumping, while the \PV\ UV line on the right is strongly affected. Adapted from \citet{2007A&A...476.1331O}.}% Oskinova+ 2007 
\label{f_oski}
\end{center}
\end{figure*}

\citet{2007A&A...476.1331O} employed an “effective opacity” concept in the formal integral for the line profile modelling of the famous O supergiant $\zeta$~Pup.
Figure~\ref{f_oski} shows that the most pronounced effect involves strong resonance lines, such as the resonance doublet \PV\ which can be reproduced by this macro-clumping approach -- without the need for extremely low \mdot\ -- resulting from an effective opacity reduction when clumps become optically thick. 

Given that \ha\ remains optically thin for O-type stars it is not affected by porosity\ and it can be reproduced simultaneously with \PV. This enabled a solution to the \PV\ problem (see e.g.\  \citealt{2013A&A...559A.130S, 2018A&A...619A..59S}).
%Surlan et al. 2013; Sundqvist \& Puls 2018

\subsection{The origin of wind clumping}
\label{sec_origin}

In the canonical view of structure formation via the LDI, clumping would be expected to develop when velocities are sufficiently large to produce shocked structures. For typical O-star winds, this occurs at half the terminal wind velocity, corresponding to roughly 1.5 stellar radii.

Various observational indications, including the existence of linear polarisation \citep[e.g.][]{2005A&A...439.1107D} %Davies et al. 2005
and radial dependent Stokes I diagnostics \citep{2006A&A...454..625P} %Puls et al. 2006
indicate that clumping already exists at very low velocities, and likely arises inside the stellar photosphere.
\citet{2009A&A...499..279C} %Cantiello et al. (2009) 
suggested that waves produced by the subsurface convection zone could lead to velocity fluctuations, and possibly density fluctuations, and could thus be the root cause for the wind clumping seen close to the stellar surface. 

Assuming the horizontal extent of the clumps to be comparable to the vertical extent in terms of the sub-photospheric pressure scale height $H_{\rm p}$, one may estimate the number of convective cells by dividing the stellar surface area by the surface area of a convective cell finding that it scales as ($R/H_{\rm P})^2$. For main-sequence O stars in the canonical mass range 20-60\,$M_{\odot}$, pressure scale heights are within the range 0.04-0.24 $R_{\odot}$, corresponding to total clump numbers of 6 $\times 10^3-6 \times 10^4$. These estimates could be tested through linear polarisation monitoring, probing wind clumping close to the wind base.

In an investigation of linear polarisation variability in WR stars, \citet{1989ApJ...347.1034R} % Robert+ 1989
revealed an anti-correlation between the terminal velocity and the observed scatter in linear polarisation. They interpreted this in terms of blobs growing or surviving more effectively in slower rather than faster winds. 
\citet{2005A&A...439.1107D} %Davies et al. (2005)
found this trend to continue into the lower temperature regime of the LBVs, whose winds are even slower. Therefore, LBVs are ideal test-beds for constraining wind clump properties -- due to their very long wind-flow times.
As \citet{2005A&A...439.1107D} %Davies et al.
found the polarisation angles of LBVs to vary irregularly with time, and optical line polarisation effects were attributed to wind inhomogeneity. 
Given the short timescale of the observed polarisation variability, Davies et al. (2007) argued that LBV winds consist of order thousands of clumps near the surface. 
For main-sequence O stars the derivation of the numbers of wind clumps and their sizes from polarimetry has not yet been feasible as very high S/N data is needed. This becomes feasible with the proposed \textit{Polstar} mission, which can reach polarization precision at the $1 \times 10^{-4}$
level due to very high SNR when the data is binned to $R \cong 30,$ as mentioned above.

\section{Large-scale structures}

\subsection{Magnetically confined winds}

As an additional complication, magnetic fields can influence these hot-star winds significantly, often leading to large-scale structures. Typically, their overall influence on the wind dynamics can be characterized by a single magnetic confinement parameter,
\begin {equation}
\eta_\ast \equiv \frac {B_{eq}^2 R_\ast^2}{\dot{M} v_\infty}
\end{equation}
which characterizes the ratio between magnetic field energy density and kinetic energy density of the wind, as defined in \cite{2002ApJ...576..413U}. %ud-Doula \& Owocki (2002).
The effects at large $\eta_\ast$ are primarily the topic of another \textit{Polstar}
white paper \citep[][]{Shultz2021}, but since such magnetic influences are also relevant
to clumping and wind asymmetry that affects mass-loss rate diagnostics, and can
appear even at low values of $\eta_\ast$,
we include a short description here.

Extensive magnetohydrodynamic (MHD) simulations show that, in general, for the stellar models with weak magnetic confinement, $\eta_\ast < 1$ field lines are stretched  into radial configuration by strong outflow very quickly on a dynamical timescale. However, even for magnetic confinement as weak as $\eta_\ast \sim 1/10$ the field can have enough influence to enhance density by diverting the wind material from higher latitudes towards the magnetic equator.

On the other hand, for stronger confinement, $\eta_\ast > 1$, the magnetic field remains closed over a limited range of latitude and height about the equatorial surface, but eventually is opened into a nearly radial configuration at large radii. Within closed loops, the flow is channeled toward loop tops into shock collisions where the material cools and becomes dense. With the stagnated material then pulled by gravity back onto the star in quite complex and variable inflow patterns. Within open field flow, the equatorial channeling leads to oblique shocks that eventually lead to a thin, dense, slowly outflowing ``disk'' at the magnetic equator. This is in concert with the ``magnetically confined wind shock'' model first forwarded by \cite{1997A&A...323..121B}.%Babel \& Montmerle (1997).
Such large scale wind structures are inferred most directly from time variability in the blueshifted absorption troughs of UV P Cygni profiles.

More recent MHD modelling shows that the wind structure and wind clumping properties change strongly with increasing wind-magnetic confinement. In particular, in strongly magnetically confined flows, the LDI leads to large-scale, shellular sheets ('pancakes') that are quite distinct from the spatially separate, small-scale clumps in non-magnetic hot-star winds \citep{2021arXiv211005302D}.

%ADU paragraphs about CIRs/DACs

\subsection{Discrete Absorption Components}

\begin{figure*}
\begin{center}
\hspace{-0.2cm}
\includegraphics[width=\textwidth]{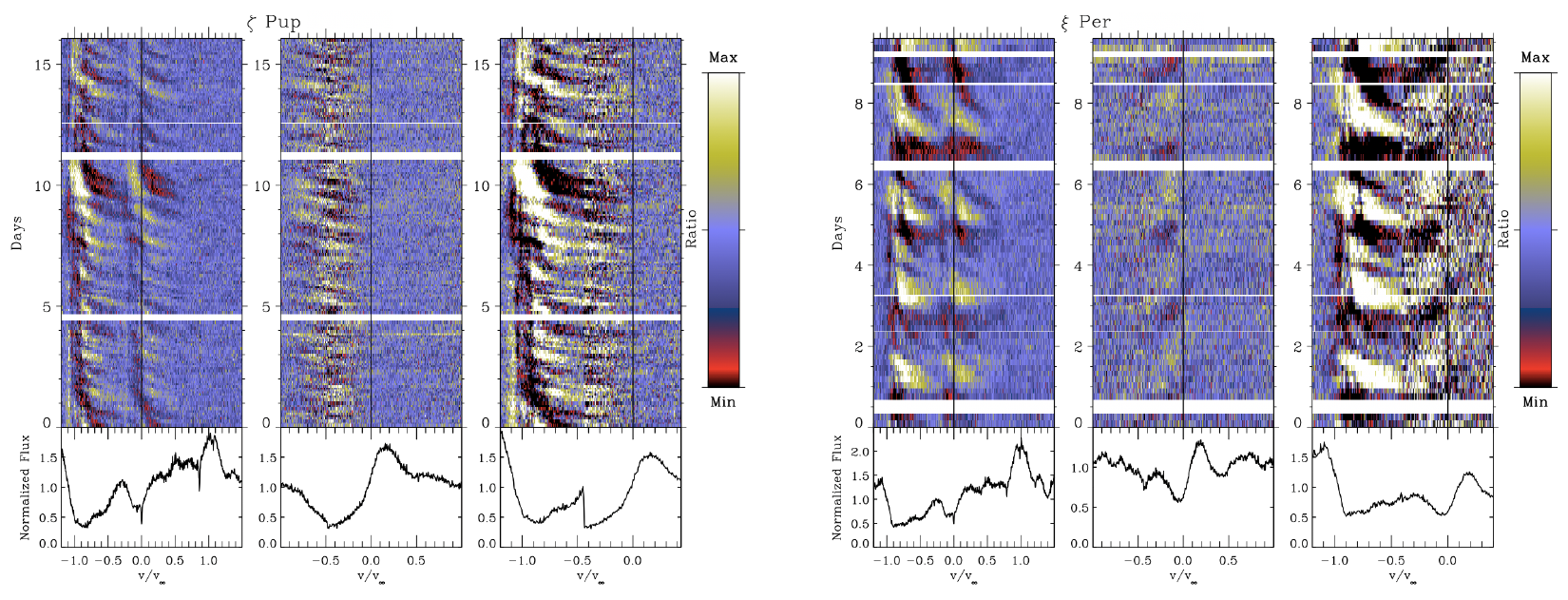}
\vspace{-0.2cm} 
\caption{ Dynamical spectra as deviations from the mean profiles shown at bottom,
taken from the IUE MEGA project for $\zeta$ Pup and $\xi$ Per.  Strong (left) and weak (center) lines are spliced together (right) to show
the continuous acceleration of the DAC features (Massa \& Prinja 2015).} 
\end{center}
\end{figure*}

Early high-resolution ultraviolet spectroscopy of massive stars obtained with the Copernicus satellite revealed the presence of sharp absorption features superimposed onto the extended blueshifted absorption troughs of far ultraviolet (FUV) resonance lines \citep{1975ApJ...199..691U,1976ApJ...203..386M,1976ApJS...32..429S}, often located near the terminal wind velocity. Initially dubbed ``narrow absorption components", or NACs, they were later found to be snapshots of a dynamic feature now known as ``discrete absorption components" (or DACs). The discovery of DACs was made possible by the advent of time-resolved high-resolution FUV spectroscopic observations obtained with the International Ultraviolet Explorer (IUE). They were found to start out as broad, shallow absorption features at low velocity, narrowing and deepening (in some cases saturating) as they migrate to higher velocities over characteristic timescales on the order of days (e.g. \citealt{1987ApJ...317..389P}). Furthermore, they were found to occur nearly ubiquitously among O stars \citep{1989ApJS...69..527H}. Since they cover a large range of velocities in the absorption trough, they must be caused by features covering a large radial extent within the stellar wind, and be indicative of a physical process universally present in massive stars.

Determining whether the variability seen in the UV time series spectra is 
due to material originating near or at the surface of the star and then 
propagating outward, or low speed material at large distances from the 
star which simply appears at low velocity is a goal that we seek to 
address with \textit{Polstar}. Should the origin of all of the features be tied to 
the stellar surface, this would have implications about the structure of 
the winds (how does the component of the wind which shapes saturated wind 
lines co-exist with large scale structure), the structure of the 
photosphere (what are causes the photospheric irregularities that give 
rise to the different wind flows), X-ray production (can some X-rays originate 
along the interfaces of 
the CIRs) and the theoretical derivations and simulations based on a 
smooth wind.

Generally, there is no way to directly determine the location in the wind where an absorption feature at a specific velocity originates. For example, a low velocity feature could be near the stellar surface, accelerating outward, or due to a slowly moving parcel of wind material far out in the wind, which is re-accelerating after being shocked. This ambiguity is always the case for resonance lines, but not for excited state lines. An excited state wind line arises from an allowed transition whose lower level is the upper level of a resonance transition (typically below 900A). They frequently appear as wind lines in O stars with strong mass fluxes. One of the most commonly observed excited state lines is N\,{\sc iv}$\lambda$1718\AA , whose lower level is the upper level of the N\,{\sc iv}$\lambda$765\AA\ resonance line. Because a strong EUV radiation field is required to populate an excited state line, these lines can only exist close to the star (this is what gives excited state lines their distinctive shapes). As a result, a feature which appears at low velocity in an excited state line must originate close to the stellar surface

Unfortunately, this same property means that excited state lines weaken quickly at large distances from the star and, as a result, rarely extend to very high velocities. This challenge can be met however by the superior S/N spectra offered by \textit{Polstar}. 
We see, therefore, how resonance and excited state lines complement one another. If a feature (indicated by excess or reduced absorption) appears at low velocity in an excited state line and then joins a high velocity feature in a resonance line (whose low velocity portion may be saturated), this provides evidence that the excess or deficiency of absorbing material which caused the feature originated close to the stellar surface (where the radiation field is intense) and then propagated outward, into the wind.

A powerful diagnostic is to compare and 'splice' together the temporal 
behaviour seen in the Si\,{\sc iv}$\lambda\lambda$1400\AA\ resonance line doublet and N\,{\sc iv}$\lambda$1718\AA\ 
excited state singlet (cite $\zeta$\,Pup and $\xi$\,Per figures). Our analysis is 
relatively simple, and based primarily on inspection of dynamic spectra. 
Dynamic spectra are images of time ordered spectra normalized by the mean 
spectrum of the series. The properties of coherent structure seen in the 
full comparison between these lines will allow is to (i) test whether 
every absorption feature observed at low/near-zero velocity connects to a 
feature at high velocity. (ii) reveal “banana” shaped patterns, which are 
the accepted signatures of CIRs tied to the stellar surface. (iii) 
establish connections between the wind features and the stellar surface. 
(iv) constrain any instances of evidence features formed (at intermediate 
velocity) in the wind: if all features can be traced back to the surface 
and they move through the profile monotonically with velocity, this would 
imply that we only see evidence for porosity in the dynamic spectra, and 
not “vorocity”. (v) test the notion that features in dynamic spectra must 
be huge, with large lateral extent in order to remain in the line of sight 
for days, persisting to very large velocity.

\subsection{Corotating Interaction Regions}

As mentioned above, one of the leading hypotheses to explain DACs involves large-scale structures extending through most of the radial extent of the wind. \citet{1986A&A...165..157M} suggested that ``corotating interaction regions" (or CIRs), as seen in the solar wind, could be responsible for the observed phenomenology. This model involves rotation of the star changing the conditions at the footpoint of flows along the line of sight to the observer, 
producing regions of either stronger or weaker wind that interact because of their
different accelerations along that line. 

One way
to achieve this quasi-periodic footpoint modulation is to adopt \textit{ad hoc} brightness spots on the stellar surface. \citet{1996ApJ...462..469C} tested both dark and bright spots using 2D hydrodynamical simulations, finding that only synthetic line profiles computed using the CIRs that emanate from bright spots reproduce the observed DAC structure. 
Also, they found that while there was a slight overdensity within the CIRs, the real source of the enhanced absorption in DACs stems from the velocity plateau (or ``kink") that arises when sparser, faster material from the bulk of the wind plows into the denser, slower material driven by the bright spots.

This model did not attempt to account for the existence of spots, although some ideas have
been suggested, including the possibility that magnetic flux tubes could be causing
depressions in the stellar surface to maintain lateral balance in the combined gas and
magnetic pressure.
These depressions could penetrate to higher temperatures and produce brightening, and
growing observational evidence suggests this may occur in a variety of massive stars
not observed to have globally strong fields (e.g., \citealt{2019MNRAS.490.2112B};
\citealt{2020NatAs...4.1092M}).
However, weak photometric variations \citep{2018MNRAS.473.5532R}
place problematical constraints on the size and
brightness of spots that could seed wind features large
enough to explain DACs, so the interpretation remains unclear. 

The combination of high spectral resolution with polarimetric sensitivity, such as
provided by \textit{Polstar}, can help
constrain the attributes of the structures responsible for DACs.
The IUE MEGA campaign well characterized DACs in the time domain for only a few
targets, and at a SNR that could only detect the largest features.
But the full nature of such features remains ambiguous when all that is available
is information about their projection against the star, and the interpretation of
the line Doppler shifts, though valuable, is
complicated by the details of the velocity structure.

\begin{figure}
    \centering
    \includegraphics[trim=10 0 50 400, clip, width=9cm]{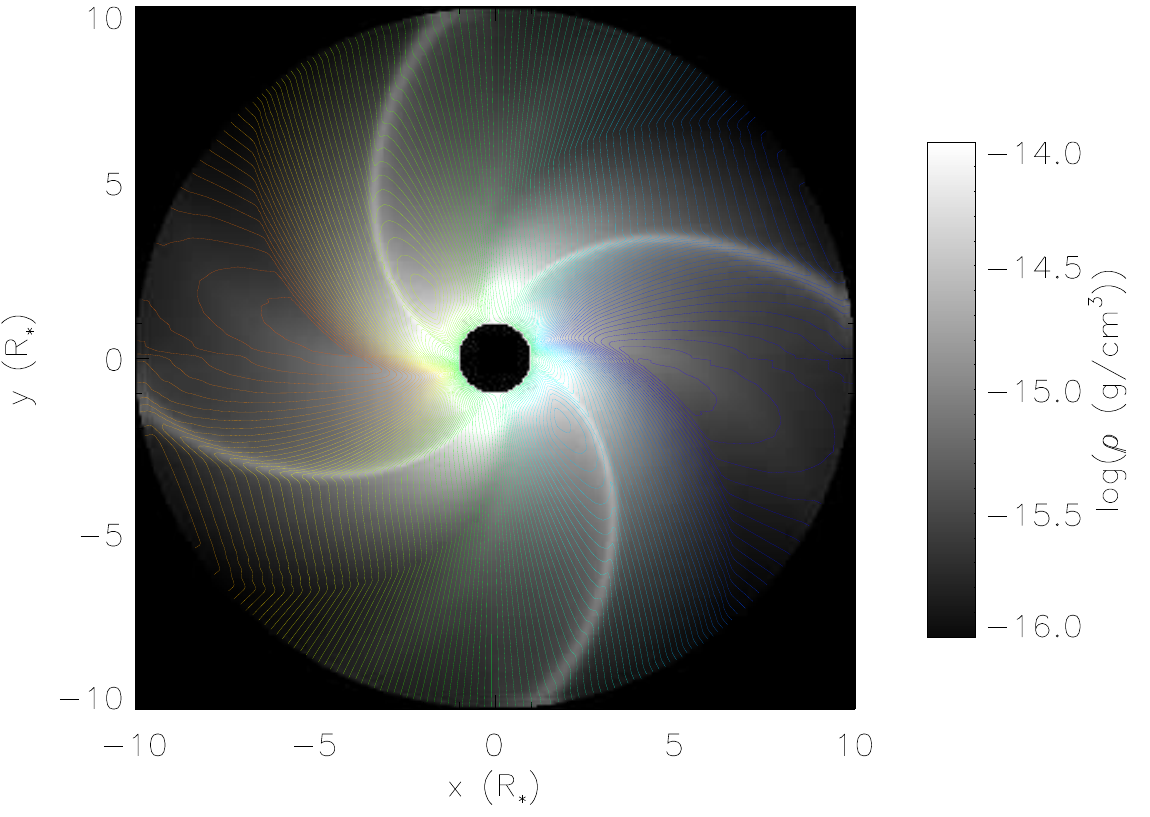}
    \caption{Adapted from \citet{2017MNRAS.470.3672D}: 2D equatorial slice of a wind with CIRs created by four equally spaced bright spots. The greyscale background represents density, and color-coded line-of-sight isovelocity contours are overlaid for an observer along the positive x-axis (blue corresponds to high-velocity blue-shifted material, and red to high-velocity red-shifted material).  A 50 km s$^{-1}$ resolution is shown, illustrating that the $\sim$10 km s$^{-1}$ resolution achievable with \textit{Polstar} will spatially sample the complex structure of the wind material very effectively.  
    This will continue to be true when the data is binned to 20 km s$^{-1}$ resolution to obtain the high
    SNR desired in this experiment.  The figure demonstrates that when large-scale structure is present, spatial resolution cannot be simply
    interpreted from the Doppler shifts in the line profile; additional geometric information from continuum polarization will be necessary
    to resolve ambiguities in the global structure.}
    \label{fig:cir}
\end{figure}

The potential for complexity is exhibited in Fig.~(\ref{fig:cir}), where shown
is the resonance zones in a CIR-type model structure.
Dynamical spectra include a wealth of information from the time domain about 
structure along the path to the stellar disk,
but polarization information from the wind
patterns as they turn
away from that disk complements these constraints.
Polarization information from Ch1 is automatically obtained with the \textit{Polstar}
dynamical spectra, and binning to low resolution increases the sensitivity dramatically.
The foreground polarization from dust in the interstellar medium will be separated
from the intrinsic signal using means  
described in more detail in the \textit{Polstar} white paper
focused on the interstellar medium signal \citep{Andersson2021}.
This involves
wavelength-dependent signatures that only require some 30 points per spectrum, rather than
the 30,000 available in the dynamical spectra, so binning by a factor of 1000 will be used
to raise the SNR from $> 400$ in Table (\ref{targettable}) to $ > 10,000,$ suitable for the
required $3 \times 10^{-4}$ polarization sensitivity target, with margin.

The DACs themselves have been well characterized at the IUE SNR, but their impact on
the inferred mass-loss rates is only significant if they represent a peak in a
hierarchy of smaller structures that IUE did not have the SNR to detect.
Also, 
High spectral resolution can provide good information about the spatial distribution of material in the wind 

While DACs repeat in the dynamic spectra of OB stars (e.g. \citealt{1996A&AS..116..257K}), they are not strictly periodic. Nevertheless, they are characterized by a timescale that is understood to be related to the rotation period of a star \citep{1988MNRAS.231P..21P}. This rotational modulation prompted the hypothesis that CIRs might be a consequence of large-scale magnetic fields on the surfaces of massive stars, as have since been detected on a subset of them \citep{2017MNRAS.465.2432G} -- interacting with the stellar wind to form ``magnetospheres", as discussed above. However, using high-quality optical spectropolarimetric data, \citet{2014MNRAS.444..429D} have placed stringent upper limits on the magnetic field strengths of a sample of 13 well-studied OB stars with extensive FUV time-resolved spectra. They conclusively demonstrated that large-scale fields could not exert sufficient dynamic influence on the wind to form CIRs. An alternative explanation involves non-radial pulsations (NRPs), but reconciling their typical periods with the DAC recurrence timescales requires complex mode superpositions \citep{1999A&A...345..172D}.  
Therefore, the prevailing hypothesis regarding the origin of putative CIRs involves the presence of spots on the stellar surface. 

%  ADD SOMETHING ABOUT POLARIZATION SIGNATURES EXPECTED FROM CIRs

\subsection{Surface spots}

Even if surface spots seem a promising cause of CIR structures, the reasons behind their presence remain unclear.
Unlike lower-mass Ap/Bp stars and chemically peculiar intermediate mass stars (such as Am stars and HgMn stars), early-type OB stars are not expected to have chemical abundance spots, as they would be continuously stripped by the wind (e.g. \citealt{1987ApJ...322..302M}). One possible mechanism by which such brightness spots can be formed involves small-scale magnetic fields. A possible consequence of convective motions in the subsurface convection zone due to the iron opacity peak (FeCZ; \citealt{2009A&A...499..279C}; \citealt{2020ApJ...902...67S}), magnetic spots can lead to brightness enhancements by locally reducing the gas pressure, and hence the optical depth. This creates a sort of ``well", allowing us to see hotter plasma, deeper inside the envelope \citep{2011A&A...534A.140C}. While such spots have yet to be detected in OB stars, weak structured fields have now been detected on several A-type stars (e.g. \citealt{2017MNRAS.472L..30P}). There have also been further theoretical advances suggesting that stars with radiative envelopes can either have large-scale fields strong enough to inhibit convection in the FeCZ, or weak,  disorganized fields generated by that convection \citep{2020ApJ...900..113J}, leading to a bimodal distribution of field strengths and thus explaining the so-called `magnetic desert' \citep{2007A&A...475.1053A}.

Regardless of their physical origin (localized magnetic fields or NRPs), bright spots have been detected on at least two archetypal massive stars: $\xi$ Per \citep{2014MNRAS.441..910R} and $\zeta$ Pup \citep{2018MNRAS.473.5532R}. These initial studies suggest that the spots are large (angular radius greater than $\sim$10\textdegree) and relatively persistent (lasting roughly tens of rotational cycles), and lead to fairly faint variability (about 10 mmag in the optical -- only detectable with space-based facilities). Assuming that these are hot spots (which means that their brightness enhancement in the ultraviolet compared to the surrounding photosphere is greater than in the optical), further hydrodynamical simulations carried out by \citet{2017MNRAS.470.3672D} showed that the spot properties inferred from the light curve of $\xi$ Per could also lead to CIRs that quantitatively reproduce the DACs observed for this star.
Clearly the UV band has much to say about the importance of these spots, and \textit{Polstar}'s observations will have a lot
to contribute to this issue, while in the process of establishing corrections to the mass-loss rates.

% ADDITIONAL THINGS TO COVER:
% \begin{itemize}
%   \item Given stronger brightness contrast in the UV, LPV in photospheric lines could help track surface spots.
%
%    \item Brightness spots can lead to linear polarization signatures
%    
%    \item Localized magnetic fields can lead to additional polarization signatures (circular+linear), 
%     though that might be difficult to detect with Polstar...
%    
% \end{itemize}

\begin{figure*}[t]
\label{figuremodelFlorian.png}
\begin{center}
\hspace{-0.2cm}
\includegraphics[width=\textwidth]{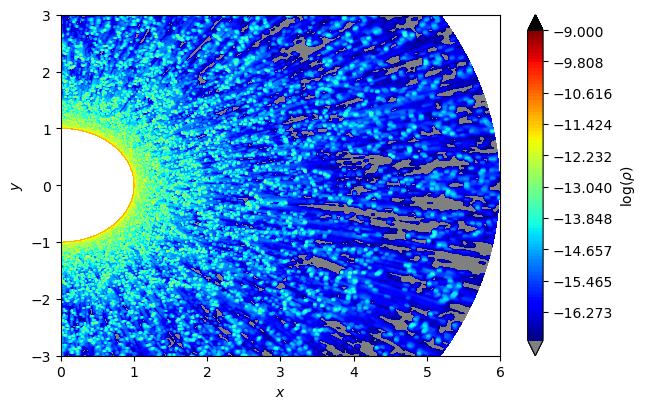}
\vspace{-0.2cm} 
\caption{Shown is the density structure in the 2D LDI simulation. To limit computing time, the radiative driving
is entirely radial.} 
\end{center}
\end{figure*}

\section{Hydrodynamic Modeling}

Recent models of the line-driven instability (LDI) using 2D hydrodynamics driven by 1D radiative transfer, simplified to allow
dynamical resolution of the resulting clumping, gives the density structure shown in Fig. (5).  %\ref{modelFlorian.png}
These models do not use any variations at the lower boundary, so the extreme degree of clumping is entirely self-excited due
to the rapid growth rate of the intrinsic instability.
The lateral scale is limited only by the tiny scale ($\cong R/100$) of the Sobolev length, producing thousands of clumps in all.
As the radial scale is stretched out by the radial instability in the velocity gradient, the clumps appear radially elongated,
which causes them to produce absorption fluctuations that are wide in velocity ($> 100$ km s$^{-1}$), but only a few percent in depth,
as can be seen in the time varying P Cygni profile in Fig.~(6). %\ref{PCygniFlorian.png}

Since the model is only 2D, there is an artificial azimuthal symmetry around the line to the observer, which
produces circularly coherent structures that overestimates the magnitude of the variances shown.
Assuming the azimuthal coherence length should actually be on the Sobolev scale, $\cong R/100$, an order of 
magnitude of cancellation in the temporal variations could be expected.
This will require further modeling to explore, but given that the largest features have a variance approaching
10\% in this azimuthally symmetric treatment, the largest temporal variations might appear at the 1\% level or less.
Nevertheless, the \textit{Polstar} 
experiment described above achieves SNR in the range 400 - 1000, depending on the target
brightness, so temporal variations over a day-long observation could easily characterize variances at the 1 \% level,
or even less.
The intended binning to 20 km s$^{-1}$ resolution will also easily resolve the model features, which typically show
a velocity width $> 100$ km~s$^{-1}$.

\begin{figure*}
\label{figurePCygniFlorian.png}
\begin{center}
\hspace{-0.2cm}
\includegraphics[width=\textwidth]{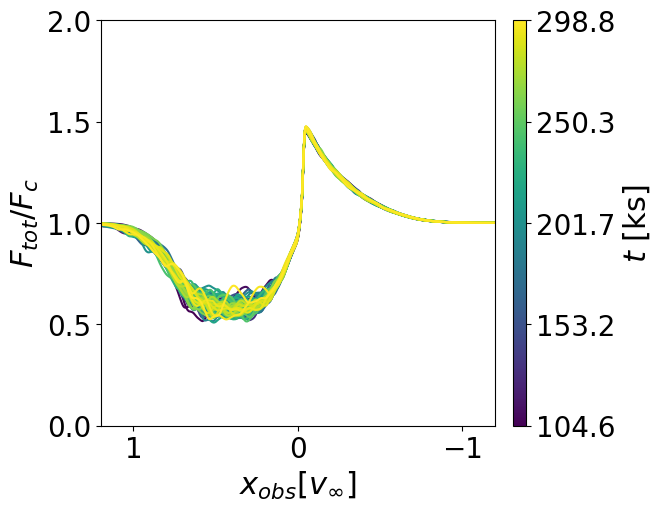}
\vspace{-0.2cm} 
\caption{Temporal snapshots of the P Cygni profile variations from the LDI simulation, showing variations of width
$> 100$ km s$^{-1}$.  The depth of the features is enhanced artificially in the 2D model by azimuthal coherence around the observer
line of sight, and could be an order of magnitude less.  The mean shape of the P Cygni absorption is less deep than a smooth model,
owing to the effects of ``vorosity".} 
\end{center}
\end{figure*}

\begin{figure*}
\label{figuredynspecFlorian.png}
\begin{center}
\hspace{-0.2cm}
\includegraphics[width=\textwidth]{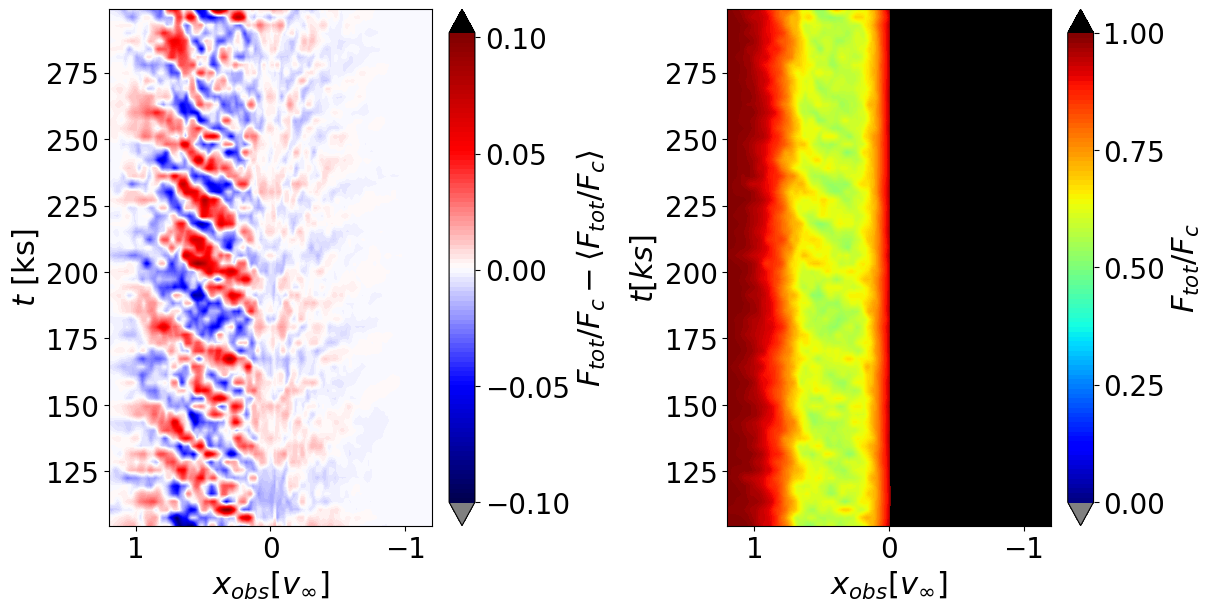}
\vspace{-0.2cm} 
\caption{Dynamical spectra from the LDI model are shown relative to the mean profile on the left, and for the full
profile on the right.  The velocity and temporal widths of the features are true to the simulation, but their magnitude
is overestimated by perhaps an order of magnitude because of the model 2D azimuthal symmetry around the observer line of sight.} 
\end{center}
\end{figure*}

The dynamical spectrum produced by this model is shown in 
Fig. (\ref{figuredynspecFlorian.png}). % (\ref{figuredynspecFlorian.png})
Again the artificial azimuthal coherence in the model may overestimate the scale of the features, but their general
attributes are seen to be highly reminiscent of observed dynamical spectra.
The model features have an acceleration that is seen to be nearly constant, and somewhat faster than
what has been so far observed (the scale of the acceleration in the simulation is $\cong$ 0.2 km s$^{-2}$,
whereas the few observed accelerations are perhaps about 1/4 that rapid).
This may be because observed features are primarily limited to large DACs that may be linked to CIRs induced by rotating surface features such as spots,
not the intrinsic wind features simulated here.

The higher SNR accessible by the 
\textit{Polstar} \textit{S2}
experiment will allow us to ascertain whether there are different acceleration
rates for features linked to surface inhomogeneities, as opposed to self-excited clumps in the wind.
If such potentially more rapid accelerations exist in the data,
the wide features that appear in the simulation shown can still be temporally 
tracked using the exposure times in this experiment by binning the data to 100 km s$^{-2}$, 
approximately doubling the already high SNR in the process.
Alternatively, if the accelerations are more at the scale of the surface gravity of the star, as may tend to be true
of the few DACs that have been well resolved, then the finer binning at the sound speed level ($\cong$ 20 km s$^{-1}$)
will match well the exposure times and produce the experiment described in section 3 above.

Also, the seeming quasi-periodicity seen in these simulations, despite including
no rotation, allows for the
possibility that the wind is ``flapping'' on its own intrinsic timescale.
These day-long timescales
might then be only coincidentally related to rotation periods of fast rotators
like zeta Puppis, and the DACs might then be unrelated to spots or CIR features.
This possibility can also be tested by the \textit{Polstar} \textit{S2} experiment,
because enough stars will be targeted to include a range of inclinations and rotation rates.
CIR-type structures would depend on those variables, but not intrinsic wind effects.

It should be noted that although these simulations include the LDI, they also
involved physically motivated boundary conditions that are conducive to surface variations
from the ``nodal topology'' 
of the line-driven critical point citep{2015MNRAS.453.3428S}.  % Sundqvist
Hence, it is not yet clear if the features seen in these simulations develop in the
wind due to the instability, or propagate from the surface due to the nonlinearity
of the driving.

Finally, the fact that observed DACs must cover a significant fraction of the stellar disk
(in order to produce such deep features) presents challenges to every one of these
models.  With an order of magnitude improvement in SNR over the IUE MEGA campaign, 
\textit{Polstar} will be able to search for a hierarchy of smaller structures that
cover less of the stellar disk, to place the more blatant DAC structures into a fuller context
of other features that will help ascertain their true nature.

\section{Evidence for Structure from Polarization}

As described above, the \textit{Polstar} mission will include polarization measurements
in support of objective \textit{S2} taken in Ch1 and binned in wavelength to increase the SNR
to the levels required to detect small polarization signals at the level of a 
few times $10^{-4}$.
The goal will be to detect free electron scattering in large-scale wind structures like CIRs,
and also time variation in the polarization position angle due to stochastic scattering
in a host of smaller scale clumps.
Also, if surface spots play a role in inducing the wind structure, especially the large-scale
DAC structures, the location on the stellar surface of these spots, when near
the limb, can be correlated with the
position angle they produce.
This is the manner in which polarization information complements the spectroscopic
signals, but for targets with a significant polarization contribution from the interstellar
medium (ISM), the ISM component will need to be removed.
This will be done using the well-understood wavelength dependence of the ISM component,
as well as its expected time independence, contrasted with the time dependence of the
intrinsic stellar component.
More details are given in the white paper describing the objectives
related to ISM polarization \citep{Andersson2021}, as mentioned previously.

\section{Conclusions}

The \textit{Polstar} \textit{S2} experiment is built to resolve and track small-scale clumping in line-driven winds at SNR $> 400$, and also to connect
larger scale structure with UV polarization signatures stemming from free-electron scattering.
The purpose is to quantify corrections of mass-loss rate determinations that are sensitive to density inhomogeneity, such as $\ha ~$
and free-free radio emission.
This will lead to more accurate determinations of how much mass high-mass main-sequence stars and blue supergiants lose, setting the
stage for, and significantly altering, later evolutionary phases for single stars and binaries alike.
In the process, the physics of the driving of these winds, and their profound importance on galactic ecology, will be better understood.

\begin{acknowledgements}

RI acknowledges funding support from a grant by the National Science Foundation, AST-2009412.
Scowen acknowledges his financial support by the NASA Goddard Space Flight Center to formulate the mission proposal for \textit{Polstar}.
Y.N. acknowledges support from the Fonds National de la Recherche Scientifique (Belgium), the European Space Agency (ESA) and the Belgian Federal Science Policy Office (BELSPO) in the framework of the PRODEX Programme (contracts linked to XMM-Newton and Gaia).
SE acknowledges the STAREX grant from the ERC Horizon 2020 research and innovation programme (grant agreement No. 833925), and the COST Action ChETEC (CA 16117) supported by COST (European Cooperation in Science and Technology). A.D.-U. is supported by NASA under award number 80GSFC21M0002.
AuD acknowledges support by NASA through Chandra Award number TM1-22001B issued by the Chandra X-ray Observatory 27 Center, which is oper- ated by the Smithsonian Astrophysical Observatory for and on behalf of NASA under contract NAS8-03060. NS acknowledges support provided by NAWA through grant number PPN/SZN/2020/1/00016/U/DRAFT/00001/U/00001.
LH, FAD, and JOS acknowledge support from the Odysseus program of the Belgian Research Foundation Flanders (FWO) under grant G0H9218N.

\end{acknowledgements}

% \bibliography{S2}{}

\end{document}